\documentclass{ws-procs975x65}

\begin{document}

\title{Are classically singular spacetimes quantum mechanically singular as well?}

\author{D. A. KONKOWSKI}

\address{Department of Mathematics, \\
U.S. Naval Academy, \\ 
Annapolis, Maryland, 21402, USA\\ 
E-mail: dak@usna.edu}

\author{T. M. HELLIWELL and V. ARNDT }

\address{Department of Physics, \\ 
Harvey Mudd College, \\
Claremont, California, 91711, USA\\
E-mail: T\_Helliwell@HMC.edu}

\maketitle

\abstracts{
Are the classical singularities of general relativistic spacetimes, normally defined by the incompleteness of classical particle paths, still singular if quantum mechanical particles are used instead? This is the question we will attempt to answer for particles obeying the quantum mechanical wave equations for scalar, null vector and spinor particles. The analysis will be restricted to certain static general relativistic spacetimes that classically contain the mildest true classical singularities, quasiregular singularities.}

\section{Introduction}

Here we extend the definition of quantum singularity to include null vector and spinor particle evolution incompleteness. We then apply the definition to study a class of spacetimes with classical quasiregular singularities. This conference proceeding is based on [\refcite{HKA}].

\section{Quasiregular Spacetimes}

A classical singularity is indicated by incomplete geodesics or incomplete paths of bounded acceleration [\refcite{HE}] in a maximal spacetime. Classical singularities have been classified by Ellis and Schmidt [\refcite{ES}] into three basic types: quasiregular, non-scalar curvature, and scalar curvature. The mildest is quasiregular and the strongest is scalar curvature. Here we are solely concerned with spacetimes with quasiregular singularities.

\par The class of quasiregular spacetimes considered possess disclinations and dislocations [\refcite{PS}]. It is a 2-parameter family of static Gal'tsov-Letelier-Tod (GLT) spacetimes [\refcite{GL}, \refcite{Tod}]:

\begin{equation}
    ds^{2} = -dt^{2} + dr^{2}+ \beta^{2}r^{2}d\phi^{2} +
    (dz + \gamma d\phi)^{2}.
    \label{eq:6}
\end{equation}

\noindent Here $\beta$ and $\gamma$ are constants. These spacetimes contain idealized cosmic strings (disclinations) for $\beta^{2} \neq 1$ and spacelike screw dislocations for $\gamma \neq 0$. 

\section{Quantum Singularities}

Horowitz and Marolf [\refcite{HM}] have defined a static spacetime as quantum mechanically singular if the spatial portion of the Klein-Gordon wave operator is not essentially self-adjoint on a $C_{0}^{\infty}$ domain in $L^{2}$, a Hilbert space of square integrable functions. In this case the evolution of the test quantum wave packet is not uniquely determined by the initial wavefunction, the spacetime metric and the manifold.

\par This definition is easily extended to Maxwell and Dirac fields [\refcite{HKA}]. We say that a spacetime is quantum mechanically singular with respect to a Maxwell or Dirac field if the spatial portion of any component of the field operator fails to be essentially self-adjoint. We take the Hilbert space to be $L^{2}$ and the original domain to be $C_{0}^{\infty}$. To test for essential self-adjointness of the spatial portion $A$ of a component of the operator we use the von Neumann criterion $A^{*}\Psi = \pm i \Psi$ and determine the number of solutions that belong to $L^{2}$ for each $i$. If the deficiency indices are $(0,0)$, so that no solutions are square integrable, then the operator is essentially self-adjoint.

\subsection{Scalar Particles}

The Klein-Gordon equation $\square \Phi = M^{2} \Phi$ can be separated in GLT spacetime [\refcite{HKA},\refcite{HK}]. Using Eq.(1) gives a quantum singularity for $\Phi$ modes with 

\begin{equation}
	-1 < \frac{m-\gamma k}{\beta} < 1
\end{equation}

\noindent where $m$ and $k$ are separation constants, $m$ being the azmuthal quantum number and $k$ the momentum.

\subsection{Null Vector Particles}

The classical source-free Maxwell equations $A^{;\nu}_{\mu;\nu}=0$ in the Lorentz gauge $A^{\mu}_{;\mu}=0$ can be separated in the GLT spacetime by taking linear combinations of modes [\refcite{HKA}]. Using Eq.(1) gives a quantum singularity for       $A^{\mu}$ modes with

\begin{equation}
	-1 < \frac{m-\gamma k}{\beta} < 1.
\end{equation}

\noindent The same as for scalar particles.

\subsection{Spinor Particles}

The Dirac equation $i \gamma^{\alpha}\Psi_{;\alpha} = m\Psi$ for spin-1/2 particles can be separated in the GLT spacetime [\refcite{HKA}]. Using Eq.(1) gives a quantum singularity for $\Psi$ modes with

\begin{equation}
	-\frac{3}{2} < \frac{m-\gamma k +1/2}{\beta} < \frac{3}{2}.
\end{equation}

\section{Conclusions}

Maxwell and Dirac as well as Klein-Gordon fields can thus be used to test spacetimes for quantum singularity. Here they give the same {\it generic} results.  For all three field types, however, there is a range of modes that do not "feel" the singularity and if only these modes are used to test for a quantum singularity one will not be "felt". This though is not the {\it generic} case.

\par Can these quantum singularities be removed? There are at least two possible ways to do so. The first places boundary conditions at the singularity [\refcite{Wald}, \refcite{KS}]. Of course, adding boundary conditions is equivalent to adding extra physical information, physical information not contained in the initial wavefunction, the metric and the manifold alone. The second way to remove the quantum singularity is to change the Hilbert space for the wavefunction from $L^{2}$ to the first Sobolev space. Ishibashi and Hosoya [\refcite{IH}] show that requiring both the wave function and the derivative of the wave function to be square integrable will remove the singularity in the Klein-Gordon case for a cosmic string.  Of course, this is not the usual quantum mechanical Hilbert space and one may not be comfortable adding this extra condition. 

\section*{Acknowledgements}
One of us (DAK) was partially funded by NSF grants PHY-9988607 and PHY-0241384 to
the U.S. Naval Academy. She also thanks Queen Mary, University of London,
where some of this work was carried out.

\end{document}